\documentclass[sigconf,nonacm,anonymous=false]{acmart}
\usepackage[nolist,nohyperlinks]{acronym}
\usepackage[disable]{todonotes}
\usepackage{hyperref}
\usepackage{pifont}

\newcounter{rec}
\definecolor{boxgray}{gray}{0.95}
\newcommand{\StaticBox}[1]{%
    \setlength{\fboxsep}{7pt} 
    \par\noindent
    \colorbox{boxgray}{%
        \parbox[b]{\columnwidth-15pt}{
            \setlength{\parindent}{0pt}%
            \setlength{\parskip}{16pt}%
            \centering
            #1%
        }%
    }%
}

\newcommand{\recommendation}[1]{{
\StaticBox{{{\textbf{(\therec)}}}~\,\textbf{#1}
}
}
\stepcounter{rec}
}

\AtBeginDocument{%
  }

\begin{document}
{
\title{Ten Recommendations for Engineering Research Software in Energy Research} 
{
\author{Stephan Ferenz}
\email{stephan.ferenz@uol.de}
\orcid{https://orcid.org/0000-0001-9523-7227}
\affiliation{%
  \institution{Carl von Ossietzky Universität Oldenburg}
  \streetaddress{Ammerländer Heerstraße 114-118}
  \city{Oldenburg}
  \country{Germany}
  \postcode{26129}
}
\affiliation{
  \institution{OFFIS - Institute for Information Technology}
  \streetaddress{Escherweg 2}
  \city{Oldenburg}
  \country{Germany}
  \postcode{26121}
}

\author{Emilie Frost}
\email{emilie.frost@uol.de}
\orcid{https://orcid.org/0000-0003-4791-2333}
\affiliation{%
  \institution{Carl von Ossietzky Universität Oldenburg}
  \streetaddress{Ammerländer Heerstraße 114-118}
  \city{Oldenburg}
  \country{Germany}
  \postcode{26129}
}
\affiliation{
  \institution{OFFIS - Institute for Information Technology}
  \streetaddress{Escherweg 2}
  \city{Oldenburg}
  \country{Germany}
  \postcode{26121}
}

\author{Rico Schrage}
\email{rico.schrage@uol.de}
\orcid{https://orcid.org/0000-0001-5339-6553}
\affiliation{%
  \institution{Carl von Ossietzky Universität Oldenburg}
  \streetaddress{Ammerländer Heerstraße 114-118}
  \city{Oldenburg}
  \country{Germany}
  \postcode{26129}
}
\affiliation{
  \institution{OFFIS - Institute for Information Technology}
  \streetaddress{Escherweg 2}
  \city{Oldenburg}
  \country{Germany}
  \postcode{26121}
}

\author{Thomas Wolgast}
\email{thomas.wolgast@uol.de}
\orcid{https://orcid.org/0000-0002-9042-9964}
\affiliation{%
  \institution{Carl von Ossietzky Universität Oldenburg}
  \streetaddress{Ammerländer Heerstraße 114-118}
  \city{Oldenburg}
  \country{Germany}
  \postcode{26129}
}
\affiliation{
  \institution{OFFIS - Institute for Information Technology}
  \streetaddress{Escherweg 2}
  \city{Oldenburg}
  \country{Germany}
  \postcode{26121}
}

\author{Inga Beyers}
\email{beyers@ifes.uni-hannover.de}
\orcid{https://orcid.org/0000-0002-9566-770X}
\affiliation{%
  \institution{Leibniz University Hanover, Institute of Electric Power Systems}
  \streetaddress{Appelstraße 9A}
  \city{Hannover}
  \country{Germany}
  \postcode{30167}
}

\author{Oliver Karras}
\email{oliver.karras@tib.eu}
\orcid{https://orcid.org/0000-0001-5336-6899}
\affiliation{%
  \institution{TIB - Leibniz Information Centre for Science and Technology}
  \streetaddress{Welfengarten 1B}
  \city{Hannover}
  \country{Germany}
  \postcode{30167}
}

\author{Oliver Werth}
\email{oliver.werth@offis.de}
\orcid{https://orcid.org/0000-0002-6767-5905}
\affiliation{
  \institution{OFFIS - Institute for Information Technology}
  \streetaddress{Escherweg 2}
  \city{Oldenburg}
  \country{Germany}
  \postcode{26121}
}

\author{Astrid Nieße}
\email{astrid.niesse@uol.de}
\orcid{https://orcid.org/0000-0003-1881-9172}
\affiliation{%
  \institution{Carl von Ossietzky Universität Oldenburg}
  \streetaddress{Ammerländer Heerstraße 114-118}
  \city{Oldenburg}
  \country{Germany}
  \postcode{26129}
}
\affiliation{
  \institution{OFFIS - Institute for Information Technology}
  \streetaddress{Escherweg 2}
  \city{Oldenburg}
  \country{Germany}
  \postcode{26121}
}
}

\renewcommand{\shortauthors}{Ferenz et al.}

\begin{abstract}

  Energy research software~(ERS) is a central cornerstone to facilitate energy research.
However, ERS is developed by researchers who, in many cases, lack formal training in software engineering. This reduces the quality of ERS, leading to limited reproducibility and reusability. 
  To address these issues, we developed ten central recommendations for the development of ERS, covering areas such as conceptualization, development, testing, and publication of ERS. 
  The recommendations are based on the outcomes of two workshops with a diverse group of energy researchers and aim to improve the awareness of research software engineering in the energy domain. The recommendations should enhance the quality of ERS and, therefore, the reproducibility of energy research.

\end{abstract}




\maketitle
}
\section{Introduction}
\label{sec:intro}

\ac{ERS} is defined as "software used in the scientific discovery process for understanding, analyzing, improving, and designing energy systems" \cite{ferenz_towards_2023}.
\ac{ERS} is used in multiple ways in the energy domain, e.g., for modeling specific components of energy systems, analyzing data, or implementing control strategies. A significant proportion of energy research is simulation-based, often requiring an interplay of different complex simulators, like grid simulation, power-plant simulation, and communication simulation. This leads to a strong reliance on (often third-party) simulation software, including (co-)simulation frameworks.
Thereby, \ac{ERS} contributes significantly to energy research and serves as the foundation for most research outcomes in this domain. Thus, there is a direct link between the quality of results in energy research and the quality of \ac{ERS}. Additionally, the quality of \ac{ERS} also determines if research is reproducible and if \ac{ERS} is reusable as emphasized by the FAIR (Findable, Accessible, Interoperable, and Reusable) principles~\cite{barker2022introducing}.

\ac{ERS} is often developed by energy researchers with diverse backgrounds, e.g., in electrical engineering, mechanical engineering, economics, or computer science. Many of them did not receive any training in software engineering. Therefore, \ac{RSE} regularly does not consider classical software engineering practices \cite{hasselbring2020fair}. Even computer scientists can not completely rely on their training since general software engineering practices are not fully transferable to research software \cite{heroux_research_2022}. 

Recommendations on software engineering can help researchers write better code and make code reusable. Such recommendations already exist in  computational biology \cite{list2017ten}.

Currently, no publication focuses generally on recommendations for \ac{RSE} or on \ac{RSE} in energy research. While some recommendations from other domains can be applied to the energy domain, \ac{ERS} has some unique characteristics, like its high focus on simulations across different expertise fields and its development close to industry. 

To close this gap, we developed recommendations for engineering \ac{ERS} for our community together with energy researchers. 
In this work, we present our ten recommendations for better \ac{ERS} engineering, starting from whether an idea is worth implementing over fundamental aspects like documentation and testing to building a community around a software project. These recommendations present a starting point for energy researchers to improve their software engineering skills and develop better \ac{ERS}. They were derived and validated from two workshops with other energy researchers. 

In \autoref{sec:chara}, we provide an overview of the typical characteristics of \ac{ERS}. Afterward, we present the review of the literature on general \ac{RSE} and \ac{RSE} in other domains, which we describe in more detail in \autoref{subsec:rel_work}. Then, we outline the method with which we derived the recommendations in \autoref{sec:method}. Our main contribution, the recommendations, are presented in \autoref{sec:guidelines}. Finally, we discuss our results in \autoref{sec:dis} and give an outlook of the further required work in \autoref{sec:outlook}.

\section{Characteristics of ERS} \label{sec:chara}

Research software, in general, has several unique characteristics. It is often highly specific and mainly created to solve one/few particular use case(s). The software is usually not designed for reuse and has a short lifecycle. Generally, it is hard to comprehend the software, even for researchers from the same domain, due to the inherent complexity of the solved problems. Further, most researchers are not explicitly trained in software engineering. Another fundamental characteristic of research software is that the software requirements are not known beforehand but often only evolve during development~\cite{hasselbring_toward_2024,felderer_investigating_2025}. 

\ac{ERS} shares these characteristics of general research software, while some common aspects can be considered more critical in energy research. Additionally, \ac{ERS} also has some special characteristics:

\begin{description}
    \item [Interdisciplinary research] Energy researchers have diverse backgrounds ranging from social science over engineering to different natural sciences and mathematics. As a broad range of domains are involved, it is hard to establish a common language. Therefore, it is difficult to achieve a mutual view of the problems that must be solved as a team. This issue influences the development of \ac{ERS} as well as its capability of being understood by different interested researchers like project partners.
    \item [Applied research] Energy research is an applied research field. Therefore, cooperation and joint projects with industry are widespread. This also has implications on the engineering of the jointly created software projects, e.g., with respect to open source or use of commercial software.
    \item [Changing levels of detail] \ac{ERS} is diverse and complex and needs to support many different analysis types and detail levels (e.g., transient vs. steady-state).
    \item [Complexity] Energy researchers often look into the detailed behavior of larger systems consisting of many components. 
    \item [Data heterogeneity] \ac{ERS} needs a lot of diverse and heterogeneous data for most simulations and often produces data different in structure, type and size, which needs to be managed \cite{zhang2018big}.
    \item [Reliability] Energy systems are critical infrastructures leading to high demands on research software reliability, specifically when conducting field tests. 
    \item [Changing time horizons] \ac{ERS} often considers diverse time horizons, diverse spatial resolutions, and complex optimization problems~\cite{ENGELAND2017600,DECAROLIS2017184}. These aspects can lead to high performance requirements.
    \item [Coupled co-simulations] As huge infrastructure systems are under analysis, there is a need to couple different types of simulation, often comprising communication simulation  \cite{vogt_survey_2018,steinbrink_cpes_2019}.
\end{description}%

We generally conclude that the most essential aspects of \ac{ERS} are the people involved and working on different detail levels, spatial resolutions, and time resolutions in a diverse and highly simulation-driven research area.
\section{Literature Review}
\label{subsec:rel_work}
Several existing publications provide recommendations or best practices on \ac{RSE}. To give an overview, we looked at the following categories: publications providing recommendations, publications on the general state of \ac{RSE}, and publications on \ac{FAIR} principles for research software and their evaluation.
Regarding recommendations for research software in general, various publications exist \cite{8887228, anzt2020environment, arvanitou2021software, balaban2021ten, baxter2012research, castell2024towards, cohen2020four, crouch2014software, eisty2022developers, eisty2018survey}. Some publications consider best practices for \ac{RSE} in specific domains, as in computational biology \cite{list2017ten}. Others focus on particular aspects of \ac{RSE}, for example, the documentation of software \cite{hermann2022documenting, lee2018ten}, specific programming languages as Python \cite{irving2021research}, sharing and reusing research software \cite{park2019research}, sustainability of research software \cite{de2019makes}, or testing research software \cite{eisty2022testing}.
The aspect of open-source development is more and more discussed as well \cite{zirkelbach2019modularization, hasselbring2020open}. 
Multiple publications apply the \ac{FAIR} principles to research software \cite{barker2022introducing, hasselbring2020fair, katz2021taking, gruenpeter2020m2}.
Additionally, several publications provide rules for software engineering in general, not focusing on research software \cite{brack2022ten, david2023task, jones2009software, sandve2013ten, schlauch2018software, stodden2016enhancing, stodden2013best, struck2018research, wagner2006creation}. Some publications also consider specific aspects in this area that might be applied to energy software engineering, such as robust software \cite{taschuk2017ten} or code quality \cite{trisovic2022large}. 
\par 
Based on our extensive research, we identified four topics frequently discussed in the literature:
\begin{description}
\item[Architecture and software design] This includes the modularity and reuse of existing software. The literature suggests considering community standards for input/output data~\cite{castell2024towards}, providing a comprehensible, extensible, modular, and easily exchangeable structure, as well as usage of design patterns and common rules regarding programming styles, the programming language \cite{drlsoftwareinitiative}, and iterative processes regarding requirements~\cite{johanson2018software}.
\item[Development process] Here, the development process itself, organization, environment, documentation, and code reviews are discussed. In literature, recommendations consider the software's reproducibility, usability, and maintainability. Additionally, specific recommendations on using version control systems \cite{castell2024towards}, repository structures following community standards \cite{castell2024towards},  continuous refactoring and testing \cite{drlsoftwareinitiative} or particular aspects, such as useful filenames \cite{trisovic2022large} exist.
\item[Testing] Test strategies and types of tests are considered. The literature recommends automated testing \cite{castell2024towards, drlsoftwareinitiative}, different testing types such as module, integration, system or acceptance tests \cite{drlsoftwareinitiative}, end-to-end testing \cite{hunter2021ten}, using test strategies, and defining and ensuring test coverages \cite{drlsoftwareinitiative}. 
\item[Publication of software] Aspects as licenses and the community are considered here. In the literature, publishing code~\cite{barnes2010publish} or developing open source \cite{hasselbring2020open} is highly recommended. Publications consider reproducibility and reusability of code \cite{gruenpeter2020m2}, recommending publishing everything necessary for reproducibility~\cite{stodden2016enhancing}. Additionally, appropriate licenses are suggested \cite{castell2024towards}, using persistent identifiers (PIDs) \cite{castell2024towards}, and providing well-structured metadata~\cite{castell2024towards}.
\end{description}

Although lots of literature discusses the general topics, none of the publications derive general recommendations for \ac{RSE} or 
consider the requirements and constraints of software engineering for \ac{ERS}. 
In contrast, we aim for a single publication that covers the most important aspects of \ac{RSE} in the energy domain, with domain-specific requirements and examples, such that the reader can gain a sufficient overview by reading a single publication.

\section{Method}
\label{sec:method}

This section gives an overview of the method used to derive our ten recommendations. To find, organize, and prioritize the recommendations, we used an exploratory approach \cite{harrison_inventor_2024}.

Generally, we focused on research software for modelling, simulation and data analytics, and proof-of-concept software \cite{hasselbring_toward_2024}. We generally exclude the development of infrastructure software 
because this software is developed based on a different set of requirements which normally puts usability by other more in the center. 

\begin{figure}[h]
  \centering
  \includegraphics[width=.7\linewidth]{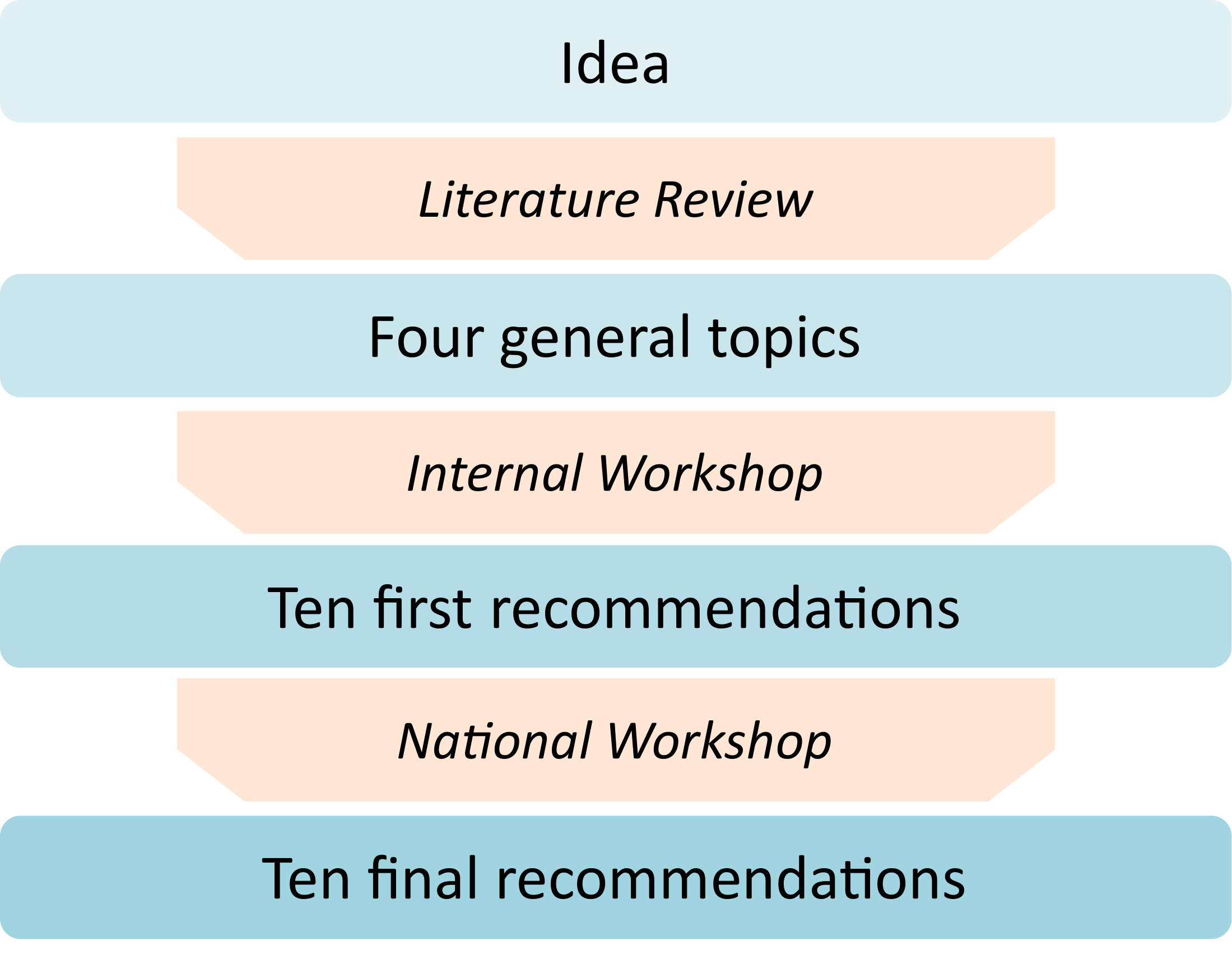}
  \caption{Flow chart of the method}
  \label{fig:flow_chart}
\end{figure}

We mainly used three sources to define the recommendations as method triangulation \cite{carter_use_2014} as shown in \autoref{fig:flow_chart}. First, we identified four topics based on the literature as described in \autoref{subsec:rel_work}. We used these topics as the foundation for an internal workshop at our institution, which we further outline in \ref{subsec:ws1}. From this workshop, we derived a first set of recommendations. We used these recommendations as a starting point for a second workshop, to which we invited multiple researchers and research software engineers from Germany. We provide more information on that workshop in \ref{subsec:ws2}. We combined the results of the second workshop with the outcomes of the first workshop to define the final recommendations presented in this paper.

\subsection{Internal Workshop}
\label{subsec:ws1}
From the related work, we identified four main topics: architecture and software design, development process, testing, and publication of software. For these topics, we picked out the key recommendations that can be applied to energy research and used them as a basis for discussion in our internal workshop. We conducted this workshop at our research institute to gather different perspectives on each topic and define a first set of recommendations.
The on-site workshop was attended by eight participants of the about 100 researchers in energy system research at our
institution\footnote{OFFIS,  \url{https://offis.de/en/applications/energy.html}, last access 2024-12-20.}
covering all energy-related research groups. The group consisted of one professor, five PhD students, and two other research staff who all have a high software engineering focus within their research groups. The workshop took place in July 2024 and was mainly held in English, while some informal discussions were also in German. The results were directly documented on boards in the room.

The workshop was structured into three main parts. The first session began with an introduction to the topic where general challenges and problems for \ac{ERS} engineering were discussed. Participants were encouraged to share their initial thoughts and ideas on the topic without bias from our preparation. Second, participants were divided into random groups and tasked with discussing the four identified topics related to \ac{ERS} engineering. Each group was moderated by one of the authors to ensure focused and productive discussions. Ideas were captured on cards, with each group contributing their insights and solutions.
In the final phase, each participant was given four points to distribute among the cards, allowing them to prioritize the most pressing challenges and innovative solutions. This voting mechanism helped identifying the key areas that require more attention and further development. 

\subsection{National Workshop}
\label{subsec:ws2}
To collect more diverse input for the recommendations and to refine 
our initial draft of recommendations, we conducted a second workshop in November 2024. We invited multiple energy researchers with high expertise in \ac{RSE}. The recruiting was based on two pillars. First, representatives of highly relevant \ac{ERS} were invited to present their software and join the workshop. Second, members of the Energy Research Center of Lower Saxony (EFZN) in Germany \footnote{\url{https://efzn.de/}, last access 2025-01-02.} were invited. The on-site workshop brought together 20 participants (excluding the authors) from 13 institutions across Germany, providing a diverse and enriching exchange of ideas and expertise. The workshop was mainly held in English, while many participants discussed in German. The results were directly documented on boards in the room.

The workshop was structured into four parts. The workshop began with a keynote on \ac{RSE} by a software engineering professor highly familiar with the topic. She provided a comprehensive overview of the field and its significance for research to lay the foundation for the further workshop. Second, specifically invited participants presented highly relevant \ac{ERS} through a poster session, facilitating an initial exchange of ideas. Additional, this brought key stakeholders to the workshop. During this session, participants were asked to identify challenges in \ac{ERS} engineering and document them. Afterward, participants were divided into four random groups and tasked with discussing the first ten recommendations. Only the headings of the recommendations, as given in \ref{subsec:ws1}, were given to the participants to allow free associations. Each group was moderated by one of the authors to ensure focused and productive discussions. Ideas were captured on cards, with each group contributing their insights and solutions. In the final phase, each participant was given ten points to distribute among the cards, allowing them to prioritize the most pressing challenges and innovative solutions. This voting mechanism helped identify the key areas that required immediate attention and further development. 

The workshop provided a collaborative platform for participants to share their experiences, brainstorm solutions, and prioritize the most critical challenges in \ac{ERS} engineering. Based on both workshops, the final recommendations were derived, which will be presented in the next section (\ref{sec:guidelines}).

\section{Recommendations}
\label{sec:guidelines}

\begin{figure*}
  \includegraphics[width=0.75\textwidth]{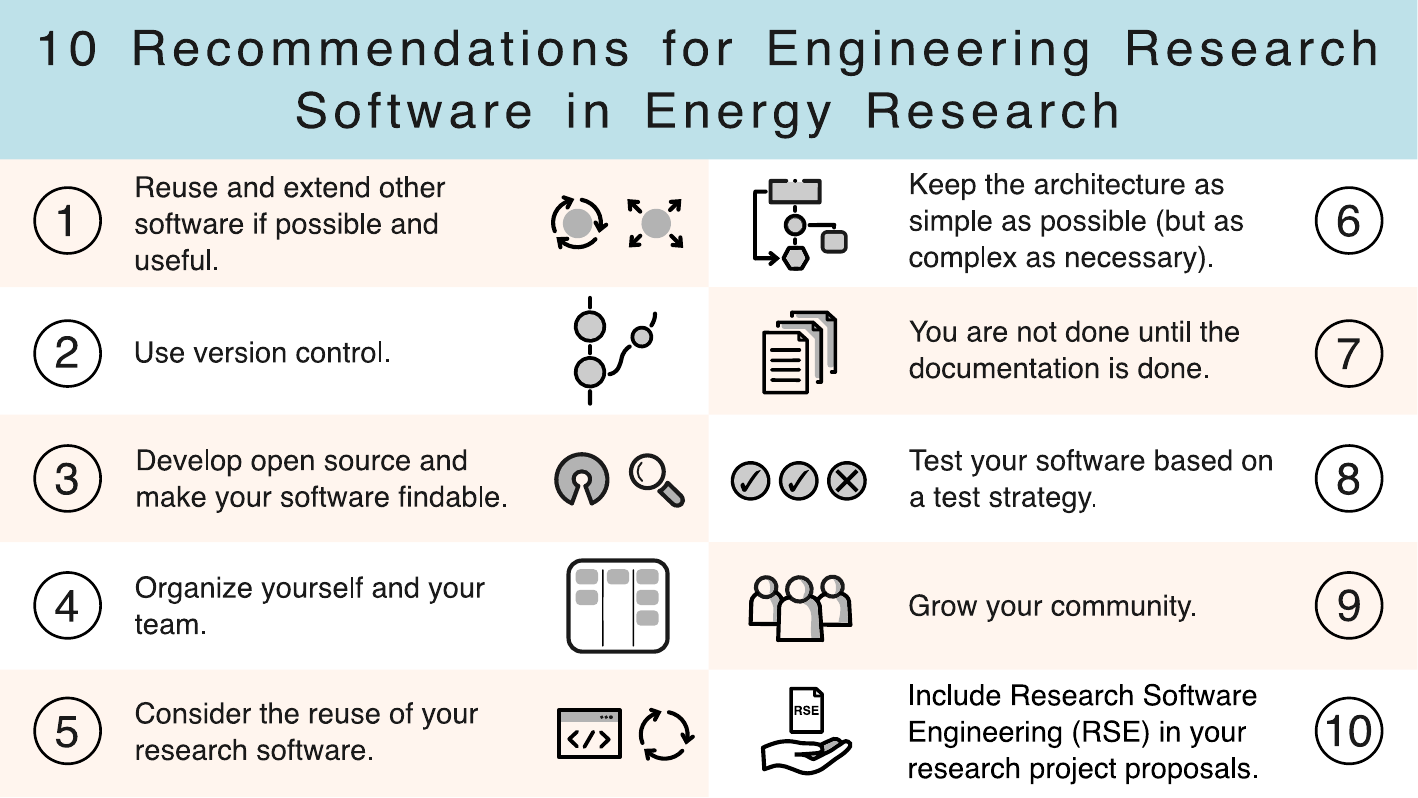}
  \caption{Overview of all ten recommendations}
    \Description{overview}
  \label{fig:over}
\end{figure*}

\autoref{fig:over} provides an overview on our recommendations for engineering \ac{ERS}. In this section, we present each recommendation in detail by providing some background on why the recommendation is important, as well as specific advice. The recommendations are ordered by required effort, so the first ones are easier to include in your daily routines than the later ones. Some of the later recommendation may also be more relevant for larger software projects or senior researchers participating in writing research project proposals. While the recommendations give first advice for each topic, we also recommend learning and collecting further information on each topic, e.g., with tutorials.

\subsection{Inventing is great, reinventing is a waste of time}
\label{sec:reuseAndExtend}

\paragraph{Background} The core of science is to build on top of other's ideas and contributions. Especially in recent years, more and more research software projects are released as open-source and can be extended and reused by the community. Still, we often feel the necessity to start from scratch and build our own library or framework. For example, a recent review identified 63 different open-source, optimization-based frameworks for energy system modeling~\cite{hoffmann_review_2024}. They all share the same underlying principle of network-based energy flow optimization with similar architectures. Many frameworks are only used by their home institutions, which indicates possible redundancy \cite{hoffmann_review_2024}. However, writing code is only one of our many diverse responsibilities.  
Overall, researchers have too little time to reimplement work already done by others. Instead, we should aim to reuse and extend existing high-quality software and focus on implementing novel research contributions. While familiarizing oneself with the documentation of other tools seems more time-consuming in the beginning, this time will be saved later on because existing code often already comes with a lot of testing, verification, and documentation which you would otherwise need to create for your new code.
This recommendation discusses creating high-quality research software without implementing everything from scratch. 

\paragraph{Recommendations} The first step of reusing software is to find existing software that solves your problem or parts of it. The natural approach is to do a literature survey beforehand. While this is an important step, not all software is published as a conference or journal publication. Instead, software registries and similar platforms can be used to find research software. For energy research,  the Open Energy Platform\footnote{\label{fn:oep}\url{https://openenergyplatform.org/}, last access 2024-12-04.} and DOE CODE\footnote{\url{https://www.osti.gov/doecode/}m last access 2025-01-20} are available. Additionally, the Helmholtz Research Software Directory\footnote{\url{https://helmholtz.software/software?&keywords=\%5B\%22Energy\%22\%5D\&page=0\&rows=12}, last access 2024-12-04.} and the \ac{JOSS}\footnote{\label{fn:joss}\url{https://joss.theoj.org}, last access 2024-12-16.} list research software without a domain focus. 
Software repositories like GitHub\footnote{\url{https://github.com/search?q=\%22power\%20system\%22\&type=repositories}, last access 2024-12-04.} can also be searched. In these cases, energy-specific keywords should be used.
The final recommendation for finding reusable software is to ask experienced colleagues and other researchers who work on similar problems. They can often tell you about hitherto unknown tools, e.g., those under development in other research projects, and which libraries are useful.
\par 
To use external software for your research project, you should check their license and see if it is compatible with your code and its envisioned use case (see also \ref{sec:openSource}). 
If the software does not entirely cover your use case, you can contact the software maintainers. They may be already discussing and planning the feature you need. If they do not want to implement it they may wish to include your software when it is finished. By explicitly communicating with other researchers, redundant software can be prevented, and as a side effect, overall collaboration in science is strengthened. This \textit{collaboration first} thinking should follow through the whole software project. For example, write issues and pull requests when finding bugs in other software you use. Not only does this help other people, but it also improves your software as a side effect.

\par 
However, there are exceptions to the rule of reusing existing software. When the existing solutions are of low quality or outdated, starting a new software from scratch can be a good idea. Also, developing an open-source alternative can have a significant scientific impact if the existing software is proprietary. 
In general, parallel development is perfectly fine to some extent. It results in competition and provides other researchers with options that have respective pros and cons.
\par 
Reusing existing research software results in higher-quality software for less effort. It accelerates scientific progress and supports overall collaboration. Hence, the following recommendation: 
\recommendation{Reuse and extend other software if possible and useful!}     

\subsection{Track the history and learn for the future}\label{sec:vcs}

\paragraph{Background} Software is constantly changing. New features are implemented, bugs are fixed, and code is improved. Sometimes, these developments also happen in parallel. For example, you are implementing a physical model to analyze the operating behavior of a battery, and you supervise a student tasked to implement a thermal model for the battery and, therefore, give them a copy of your code. When the student finishes their implementation after a couple of months, your own code has evolved, and manually integrating the master student’s code back into your own becomes a messy task of copy, paste, break, fix, test, and retest. This, and many other problems associated with changing code, multiple versions, and coding as a team, can be avoided with version control. While using version control is standard practice in software development, it is not always the case in research software. Tracking different versions of your code enables you and others to keep an overview of which feature was developed and to understand the history of the code, thereby increasing its maintainability. It also allows for the undoing of specific changes when a problem occurs. Additionally, version control enables reproducibility of the research conducted with the code by providing specific links to use  exactly the same version~\cite{sandve2013ten}.

\paragraph{Recommendations} Version control systems can help you track different software versions. The most popular version control system is Git\footnote{\url{https://git-scm.com/}, last access 2024-12-17.}.
The platforms GitHub\footnote{\url{https://github.com/}, last access 2024-12-17.} and GitLab\footnote{\url{https://about.gitlab.com/}, last access 2024-12-18} are based on Git and extend it by a user interface providing a good overview and additional features. If your institution does not allow you to use GitHub or the public GitLab, you can often also use a GitLab instance offered by your institution.

Within Git, changes to the software are included as a \textit{commit}. By adding meaningful messages to each commit, the changes are documented. At least one commit at the end of each day is recommended. 

Version control systems also allow different \textit{branches}, which are parallel versions of the same software. It is helpful to always have a working software version in one branch (usually called \textit{master} or \textit{main}). Also, this way, different developers can work in parallel without directly causing conflicts. A branch can be merged into another by a \textit{pull} or \textit{merge request}. These requests are another good option for additional comments and documentation to make the changes more understandable. In software projects with multiple developers, this is also the step where code reviews should be conducted for quality assurance.

The version control platforms also have additional features that make it easier to organize the software development. \textit{Issues} can be used to track bugs, code contributions, and new potential feature ideas. They also allow others to get an overview of what is planned next. Additionally, they provide options for \ac{CICD}, which is the practice of automating the integration, testing, and deployment of code changes to ensure reliable and efficient software delivery. In the context of research software, \ac{CICD} is especially used for automated testing of each commit (see also \ref{sec:testing}).

\par 
Version control systems are necessary to keep track of software versions while also allowing to organize the software development to a certain extent. If you are not familiar with version control, we recommend taking some basic training on it, e.g., by the software carpentry\footnote{\url{https://swcarpentry.github.io/git-novice/index.html}, last access 2024-12-17}.  

\recommendation{Use version control!}
\subsection{Research should be publicly available, and your software is part of it}\label{sec:openSource}

\paragraph{Background} Only openly available research software allows other researchers to reproduce your research results. Since research software is mainly publicly funded, it is also fair to make it publicly available. This also enables others to reuse the software, providing additional benefits like further testing of the software and contributing by creating extensions and compatible software. Besides making the software open source, it is also important to register it so other researchers can find it. These aspects are essential to make a research software \ac{FAIR}~\cite{barker2022introducing}. 

\paragraph{Recommendations} We recommend developing your research software open source. By directly starting the development as open source, challenges can be avoided when switching to open source later. Further, it allows other researchers to use your software and actively contribute as early as possible, improving your software's quality and encouraging cooperation. To develop the software open source, a suitable license is required. By choosing a license early each time another software is included, the compatibility of licenses can be checked directly. Some open source licenses are incompatible since they put certain conditions on the reuse of the code, which can conflict between licenses \cite{cui_empirical_2023}.
When choosing a license, you should check your institution's policy, which often already proposes a specific license. Also, various guides help to choose a license\footnote{e.g., \url{https://choosealicense.com/}, last access 2024-12-17.}.

Cooperation and joint research projects with industry are common in energy research. Sometimes, industry partners are critical of developing open source because they want to keep their intellectual property. Generally, we recommend discussing this topic as early as possible to find reasonable compromises. Often, certain parts can be developed open source while others remain closed source. Since open source is open to all researchers and all industries, it can also improve the exchange between industry and research. 

When you develop open source, there are specific approaches to make your code more easily reusable by others. First, the repository should follow programming-language-specific best practices. This can be achieved by starting with templates for the repository\footnote{e.g., cookie-cutter templates at \url{https://cookiecutter.io/templates}, last access 2024-12-17}. Additionally, citing your software can be made easy by including a \ac{CFF} file\footnote{\url{https://citation-file-format.github.io}, last access 2024-12-17.} in your repository
 \cite{druskat_citation_2021}.

Within the repository, it should be indicated if the software will be maintained, if support is available and if the developers are open to joining research projects, including the software. We recommend using the features of the software platform (GitHub/GitLab) to interact with potential users, e.g., by using issues.

Besides making the source code available, the software should also be findable. Therefore, we recommend registering the software in a domain-specific registry like the Open Energy Platform\footref{fn:oep}. Getting a DOI for the software is also helpful, e.g., by archiving versions of the software on Zenodo\footnote{\url{https://zenodo.org/}, last access 2024-12-17.} \cite{zenodo}. 

\par
Energy research is highly interdisciplinary \cite{tijssen_quantitative_1992}.
Therefore, we recommend adding a very general description to your software, allowing all researchers to understand its goals. This way, more researchers can identify whether the software is useful for them, increasing its reusability. GitHub also allows you to provide keywords to your repository, which again improves findability.

\par 
Platforms like GitHub and GitLab make it easy to publish your software under an open source license. Open source research software enables reproducibility and allows reusability, which can also improve your code quality. Therefore, we recommend: 

\recommendation{Develop open source \& make your software findable!}
\subsection{Being organized boosts your software project}
\label{sec:workProcess}

\paragraph{Background} This recommendation discusses how research software development teams should organize themselves.

The most important aspect here is that there is no one-size-fits-all solution.
A one-person PhD software project should be organized differently than a large team developing software with thousands of potential users.

\paragraph{Recommendations} The main recommendation is to explicitly define the software project's scope and purpose early in the process. What is the software's general use case (see \ref{sec:architecture})? How many users are envisioned (see \ref{sec:community})? 
Is the software intended for research only or for usage by industry as well? What is its expected lifespan? Will complete reimplementation be an option, e.g., for going commercial?

Also, you should get a good understanding of the available team to create the software. 
How many people are working on the software project permanently? What are their skills? 

Explicit answers to these questions are required to make wise decisions about organizing the software project and the team. 
For example, defining the software's scope helps to decide what to implement and what not. 
The team size, on the other hand, is the most important factor for defining the development process. Do we need code reviews? Do we need people responsible for specific tasks like testing, documentation? Especially, large teams must define and enforce the process explicitly to not end in chaos, like outdated documentation (see \ref{sec:documentation}) or failing pipelines (see \ref{sec:testing}).

Generally, you can think of software development organization as a scale from a high degree of agility (e.g., extreme programming~\cite{796139}) to a high degree of planning (e.g., waterfall model \cite{Petersen835760}). Consider which degree is desirable for your project, depending on the answers to the precious questions. Most research projects have a high degree of uncertainty, resulting in frequent changes in software requirements. This would naturally suit a more agile organization.

Another critical question is the developers' software engineering experience. Especially in energy research, the skill levels vary significantly, as we pointed out in \autoref{sec:intro}. 
We researchers are not expected to have the skills of a full-time professional developer. However, we \textbf{are} software developers, which mandates certain minimum skill requirements. Therefore, we should perform a skill assessment of all team members and actively provide training to fill the gaps. 
\par 
Further, it is important to assign clear responsibilities. Primarily the software project lead should be clear. It should be someone who actively participates in the software project and does not have too many other responsibilities. For example, an experienced research assistant is often a better choice than a professor who does not have the time to participate actively.

\par 
Unorganized software projects result in low-quality software, so we should explicitly discuss and define our work process. Since there are no general rules, we need clear leadership and a clear vision of our software project's scope and purpose.

\recommendation{Organize yourself and your team!} 

\subsection{Building upon research requires reusable software}\label{sec:reuse}

\paragraph{Background} To effectively build upon research, it is often required to build upon research software, and therefore, a software's reusability is a key factor to its scientific success. Research based on openly available and reusable software is cited more often than research where the software is not reusable. To achieve this, it is necessary to consider your software's applicability to more general problems (than your own). This is decided by your software's abstraction level (generalized vs. specific implementation). However, there is a conflict regarding this abstraction level most of the time. There are two extreme ends to this aspect:
\begin{enumerate}
    \item Implementing highly specialized features that serve exactly one use case, e.g., a specific heat pump model that does not enable any parameterization (e.g., to implement different coefficients of performance, capacities, or others). 
    \item Finding an abstraction for every feature, which provides maximal freedom for analyzing other use cases than the one that shall be implemented in the first case, e.g., a generalized heat pump model, which does not even have a specific mathematical model but enables you to provide an arbitrary one by yourself.
\end{enumerate}
Both ends are extreme, and usually, you want to avoid both cases. However, there is a large spectrum of possibilities between these extremes. 

\paragraph{Recommendations} We recommend finding a sensible abstraction level, as it is crucial for the reusability of the software, and good abstractions enable other researchers (including yourself and your group) to build upon your software. At the same time, no unnecessary complexities should be introduced by choosing an overgeneralized abstraction. This might render your software less usable due to the lack of specific implementations and understandability. Consequently, users need to invest more effort, and reusing becomes difficult. 

Another aspect of this is to think about the scope and objective of your software. These determine to which level your software should be abstracted. As a rule of thumb, every artifact should focus on precisely one type of research. This can be on different, relatively macroscopic levels of energy research. For example, you can develop one type of energy component model (i.e., for heat pumps) or one software to simulate decentralized individual behaviors of actors in energy communities using agent-based modeling.

Besides the level of your implementation, meta aspects, like the name of the software, can also impact the reusability of your software. Also, we recommend considering the scientific community you contribute to and how they organize themselves in open software projects. Integrating your contribution in the form of software to these software communities increases your visibility and strengthens the community as a whole. Example communities for \ac{ERS} could be based on research frameworks, e.g., oemof\footnote{\url{https://oemof.org}, last access 2024-12-18.}, or generally communities of your favorite programming language, e.g., the Julia communities\footnote{\url{https://julialang.org/community/organizations/}, last access 2024-12-18}. Furthermore, considering the target audience, which will eventually reuse your software, is essential. To improve the compatibility of the software, you can use ontologies, such as the open energy ontology\footnote{\url{https://openenergyplatform.org/ontology/}, last access 2024-12-18} to develop and validate whether the keywords used in your software and documentation fit those of the broader community.

In short, software reusability is essential for making energy research sustainable and interoperable, not only for you and your working group but also for the broader research community you are working in.

\recommendation{Consider the reuse of your research software!}

\subsection{The structure of your software determines its research potential}\label{sec:architecture}

\paragraph{Background} Choosing and developing a suitable architecture is essential to developing research software efficiently. A typical definition of software architecture would be: "The software architecture of a program or computing system is the structure or structures of the system, which comprise software components, the externally visible properties of those components, and the relationships among them." \cite{BASSSoftwareArch}. Here, we mainly think of the composition of software components, the chosen abstractions of the real world, and the applied design patterns. Inadequate and/or outdated architectures lead to different problems. They slow the development of additional features and increase the complexity of finding bugs. This happens because such architectures do not adequately induce a structure that helps you to handle the code (and its bugs) and, therefore, to extend and reiterate it. For example, using clearly defined interfaces to modules of your code can drastically improve your ability to provide different implementations for the same task (e.g., using different MILP solvers). Considering these types of decisions early saves a lot of code refactoring.

\paragraph{Recommendations} We recommend creating the most straightforward possible architecture for \ac{ERS} while making it as complex as necessary to fulfill the specific needs of your research questions. For this, looking at typical architectures and patterns to fulfill different needs is necessary. In \ac{ERS}, we often encounter two types of overall architectures: (1) single-process input-output simulations or optimizations and (2) agent-based architectures. However, when looking at the details, there are a lot of architectural design patterns to structure these types according to specific needs (i.e., Gangs of Four \cite{GoFArch}). Further, one goal of implementing an architecture is to keep it maintainable, which includes some enforcement of the architectural patterns for future development. If it is sensible that some technical data is not directly accessible, technically enforce it by restricting access to specific data fields.

To determine how to implement the architecture, it is often helpful to look at the interfaces of your components to each other (which data belongs to which component, how much data should be shared, and how the data should be shared). We recommend using standards wherever possible, for example, the \ac{CIM} \cite{uslar2012common} for power grids or the \ac{FMI}\footnote{\url{https://fmi-standard.org/}, last access 2025-01-02} for dynamic simulation models. The same is true for reusing established APIs. For example, most energy-related reinforcement learning frameworks \cite{Wolgast_OPF-Gym, hou2024rladnhighperformancedeepreinforcement, NEURIPS2023_ba748557} reuse the well-known \texttt{Gymnasium} API to be compatible with open-source algorithms.

Additionally, it makes sense to think about the interface other researchers should use when applying your software. Your architecture should allow the development of user-centric APIs (designed to be used easily, not implemented easily).

When developing \ac{ERS}, you often decide on many specific architectural aspects. Be aware of them, and document them and their reasoning. For example, if you are developing a grid model and choose to represent it as a table vs. using a graph, be aware of its advantages and disadvantages. Also, do not hesitate to change these decisions and reiterate the architecture if your software outgrows your original decision and is starting to hold it back.

Lastly, actively consider knowledge transfer for your software's lifecycle. This includes transferring your knowledge to others (e.g., using architectural documentation) and learning about basic architectural knowledge \cite{BASSSoftwareArch} and design patterns, such as Gangs of Four~\cite{GoFArch}. 

To conclude, we recommend thinking and iterating on your architecture as much as needed while aiming to keep it as simple as possible to ensure the maintainability of your software.

\recommendation{Keep the architecture as simple as possible (but as complex as necessary)!}
\subsection{Undocumented code can not be used by others, and is therefore near-to useless}\label{sec:documentation}

\paragraph{Background} When writing code, documentation is essential to help you and others understand the code. 
Documentation also helps to ensure that the code is doing what it is supposed to do. Furthermore, it allows the extension and adaption of the code and is fundamental to ensure reproducibility and reusability for others and yourself.\par

\paragraph{Recommendations} We recommend always writing documentation, no matter how large the software project is (even if only one person is involved). Any small documentation is always better than no documentation. However, software papers do not count as documentation since they cannot reflect the ever-changing nature of software.  \par
Multiple types of documentation exist:
\begin{enumerate}
    \item Documentation in the code, e.g., code docstrings, describing methods, classes, or modules. This helps others understand your code.
    \item A README file, which introduces and describes the software. This is necessary to explain the purpose and concept of the software.
    \item A documentation page, like \textit{readthedocs}\footnote{\url{https://about.readthedocs.com/}, last access 2025-01-17}, containing further descriptions of the software, helping others to fully understand the code and use it.
    \item Documentation of tutorials or concrete examples, e.g., included in the documentation page, helps others to easily apply the code.
\end{enumerate}

We recommend to consider your open source strategy (see \ref{sec:openSource}) and try to deduce the applicable documentation style and technique. If it is unclear how the documentation will be processed, you can stick to easy documentation modes. In any case, at least the following should be provided: a README file, including a summary of the purpose of the software, an installation guide, and an example script using your software.

Documentation requires considerable effort and should thus be acknowledged as part of the software development process, e.g., as an own contribution to the respective research project, and relevant resources should be allocated. It is important not to neglect it due to missing resources later on.
When planning the documentation, it is recommended to take into account the different types of documentation and the respective target audience. You can define minimal requirements for your documentation (as requirements and dependencies for the implementation, environment, and decisions made in the development process). Also, we recommend considering tests as part of the documentation, as these are necessary for others to understand your code. To support this understanding, example scenarios can be introduced in the documentation. However, all example scenarios should be fully executable.
To simplify the documentation process, we recommend combining the coding with the documentation (for example, via commit messages, see \ref{sec:vcs}) by writing readable code, including comments. Additionally, you can use templates and standards for your documentation. Many tools are available to make the documentation process easier and to help make your documentation findable. For Python, \textit{Sphinx}\footnote{\url{https://docs.readthedocs.io/en/stable/intro/sphinx.html}, last access 2025-01-17} can be used to create technical documentation, as can \textit{Documenter.jl} for Julia\footnote{\url{https://github.com/JuliaDocs/Documenter.jl}, last access 2025-01-17}.\par 

It also helps to look for best practices. These are for example provided by Github\footnote{\url{https://google.github.io/styleguide/docguide/best_practices.html}, last access 2025-01-17}, but can also be part of recommendations for your respective programming language, as \textit{pep8} for python\footnote{\url{https://peps.python.org/pep-0008/}, last access 2025-01-17}.

Due to the interdisciplinary character of energy systems research, it is recommended that the concept behind the software be explicitly discussed. People from different domains should understand the purpose of your software. Thus, also the concepts and ideas of your software, not only the software itself, should be documented. Additionally, the code itself should be documented so that other people can work with it or even extend it. 
\recommendation{You are not done until the documentation is done!}

\subsection{Testing your software means increasing trust in your research!}\label{sec:testing}

\paragraph{Background} Appropriate testing ensures your software works correctly. With testing, bugs can be found, and the software's functionality can be guaranteed. Especially when software is constantly changed, testing can help to ensure that changes do not break functionalities. 

You are likely already intuitively testing your software, for example, by using sanity checks on the software output, e.g., expecting values in a particular order of magnitude (MWh instead of kWh) or making sure that the efficiency is <1 or checking the consistency of results with conservation equations, such as conservation of mass, energy, or momentum. Other examples include adding \texttt{assert} statements to, e.g., make sure that input data has a specific structure, testing a function with dummy data before integrating it into a larger script, and exploring edge cases, e.g., ``how will my model handle a mass flow rate of zero?''.

You are likely performing these tests manually on an ``as needed'' basis. While this is efficient practice during the development phase, manual testing typically leads to a more extensive testing effort in the long run~\cite{patrick_software_2016}, and ad hoc, intuitive testing by the developer can have blind spots~\cite{mischke_automated_2022}. In addition, automated testing provides benefits such as the ability to make changes quickly and securely and perform major refactorings without the entire software immediately losing several levels of quality.

\paragraph{Recommendations} 
We recommend you elevate your intuitive, manual testing to a systematic, automated testing procedure, for example, as described in~\cite{mischke_automated_2022}. 

We recommend considering a test strategy and testing your software based on it. A test strategy describes the testing approach, including what aspects, methods, and techniques of the software are to be tested~\cite{drlsoftwareinitiative}. It should be identified early and specify how the software's testing process is designed \cite{drlsoftwareinitiative}. 
Using a strategy helps ensure that the code has been extensively and thoroughly tested, including all relevant functions. \par

In your strategy, we recommend determining what exactly needs to be tested (properties and behavior) and defining minimal requirements for that (e.g., test coverage, which defines the degree to which the code is checked through tests~\cite{drlsoftwareinitiative}). Requirements defined at the beginning of the software project should be mapped to the tests and thus checked for functionality. Similar to documentation (see \ref{sec:documentation}), tests should be considered as a contribution to the software project and should be planned accordingly. It is important not to neglect testing due to missing resources later on.
While testing, the reusability and reproducibility of the code should be considered. Thus, we recommend providing data for testing and information such as API versions, dependencies, and comparability to other software if applicable. 
Templates and standards, e.g., use testing packages for your respective type of test and programming language, as \textit{pytest}\footnote{\url{https://docs.pytest.org/en/stable/}, last access 2025-01-17} for Python, can help you when writing your tests.

We recommend taking into account different types of tests: unit, integration, and end-to-end tests. Unit tests are designed to test the functionality of specific components or modules, their restrictions and constraints, while integration tests focus on the interfaces and interaction of components~\cite{drlsoftwareinitiative}. For example, a unit test could test the performance of a photovoltaic system under different solar irradiance conditions and, thus, the functionality of the power calculation function. Integration tests would, e.g., focus on interactions between a controller of the plant and the plant itself, such as sending and responding to control signals. End-to-end tests consider the software as a whole, which helps to ensure that the overall software behaves as expected, considering the data flow over all tasks. An example of end-to-end testing is testing the entire system, including all controllers and components, from an end-user perspective. Ideally, all testing is completely automated using the \ac{CICD} capabilities of GitHub/GitLab. This way, you can ensure that every single version of your software is tested. \par

In the energy domain, software is sometimes used by people who did not develop it (people without IT knowledge). Therefore, we recommend having the software tested by potential users who were not involved in its development. By involving potential users early in the development process and having them test the installation, it is possible to ensure that the software can be used by the people it is intended for. Testing is essential if the software will be used in the field. Since we are considering software for a safety-critical system, well-designed testing is significant in energy research.
\recommendation{Test your software based on a test strategy!}
\subsection{Your software is useful to others only if they know about it}
\label{sec:community}

\paragraph{Background} In the previous recommendations, we discussed various points on how to build high-quality reusable software. However, software quality is meaningless if the community is unaware that the software project exists.
Having external users increases the scientific impact of the software and improves its quality by making it more thoroughly tested with various scenarios and edge cases. Finally, every user of your software is a potential future collaborator.
While \ref{sec:openSource} already discusses the first points as publishing the code open source and registering it in a registry, there are more ways to get attention to your \ac{ERS}.
Therefore, we will provide recommendations on increasing the number of users and overall reach of your \ac{ERS}.
\par 

\paragraph{Recommendations} The natural step to increase the visibility of your software is to publish a paper about it. While we discussed in \ref{sec:documentation} that software papers are not a replacement for documentation, they are great for increasing visibility and motivating your software. Consequently, a software paper should not contain too much technical details from the documentation. Instead, it should explain why the software is important, discuss its high-level capabilities, and convey its scope and vision. We recommend \ac{JOSS}\footref{fn:joss} \cite{jossJournal} as a modern software journal. If you want to discuss domain-specific aspects in more detail, it can also be helpful to publish your software in a conventional domain-specific journal, e.g., as it was done for \textit{pandapower} \cite{pandapower2018} and \textit{renewables.ninja} \cite{pfenninger_long-term_2016}.

\par 
The second essential step is the active engagement with the community.
Especially in the early stage of the software project, it is a good approach to contact domain experts and actively collect early feedback. First, this makes them aware of your software. Second, it prevents the developed tool from missing the needs of the researchers working in the field. Later in the process, you can actively present your software project at a conference, e.g., at the "Open Source Modelling and Simulation of Energy Systems (OSMSES)"\footnote{\url{https://www.osmses2024.org/home}, last access 2025-01-02} conference. 
\par 
If your software project uses GitHub/GitLab, their issue system is a great way to receive feedback of any kind. Feature requests provide direct information about what the users need, and questions about your software indicate where the documentation can be improved. 
In general, the software repository should be structured in a way that facilitates engagement with the community. Create a \texttt{CONTRIBUTING} file to communicate to the community how they can contribute to the software project, including, for example, how to report bugs, style recommendations, or the code of conduct.
Contact information should always be available if external researchers want to contact the maintainers. 

\par 
In summary, when developing reusable software, you should care about users and actively engage with them to improve software quality and maximize scientific impact.  

\recommendation{Grow your community!} 

\subsection{Good software development needs resources}\label{sec:proposal}

\paragraph{Background} Energy research is mainly funded through research projects. Since software engineering is often a big part of research projects it should also be covered in project proposals. Including software engineering aspects in proposals enables researchers working on research projects to develop better research software. Many of the aspects already mentioned can be included in a research project proposal.

\paragraph{Recommendations} Within a proposal, we recommend emphasizing reusing existing frameworks instead of developing new solutions as introduced in \ref{sec:reuseAndExtend}. Even if the reuse of existing software is planned, it is recommended to include some resources for maintaining and adapting the existing software.

Sometimes, it is still necessary to develop new software. In these cases, libraries that can be easily integrated into other software are better suited than complicated frameworks.
Additionally, we recommend
to formulate an overall software strategy, including a software management plan (see \cite{jackson_checklist_2018}). This should include the different development aspects during the research project, e.g., testing. Also, this strategy should cover how the software can be reused after the project. Sometimes, other research groups can even write letters of interest for reusing the software. As discussed in \ref{sec:reuseAndExtend}, we recommend choosing the licenses early, which can already be part of the proposal. 

Especially in energy research, research projects often include cooperation with industry partners. In these cases, we recommend finding a common solution for licensing with all involved partners within the research project proposal. In such cases, it can also be helpful to include resources for legal aspects of the software in the research project proposal.

When new software is developed, relevant resources should be included in the research project proposal. Besides sufficient resources for the general development of the software, including essential steps such as testing and documenting the software, this should also take usability aspects into account since the software should be reusable for other researchers. To achieve reuse, planning  resources for community interaction as described in \ref{sec:community} is also helpful.
Since most researchers are not trained in \ac{RSE}, resources for training and networking with other research software engineers are recommended. 

\par 
Software development plays an important role in research projects. By including different \ac{RSE} aspects in the research project proposal, it becomes easier to cover software engineering aspects and follow our recommendations within a research project.

\recommendation{Include \ac{RSE} in your research project proposals!}

\section{Discussion}
\label{sec:dis}

When comparing our recommendations with similar recommendations from other domains \cite{list2017ten,osborne_ten_2014,taschuk2017ten,balaban2021ten,prlic_ten_2012,hunter2021ten,wilson_best_2014, jimenez_four_2017} the recommendation on research proposal (see \ref{sec:proposal}) and on the organization of the software development (see \ref{sec:workProcess}) are new topics not covered before. Our other recommendations are included similarly in other works, although the provided details differ. The previous works less consider collaboration with industry and working in interdisciplinary teams. Reusing existing code is the only topic covered by nearly every other recommendation.

As with all studies, our method also has some limitations that must be considered. 
First, we decided to do the second workshop in-person in Germany and to use our good contacts with German researchers. While this excluded the international perspective, it gave us a more personal atmosphere with fewer technical delays, improved engagement, and more fluent in-depth discussions.
Second, energy research is highly diverse and interdisciplinary. While we covered a wide range of backgrounds, we may not have been able to include all possible perspectives in our workshops. 
Third, the diversity of possible \ac{ERS} projects results in not all recommendations being fully applicable for every \ac{ERS} development. As discussed in \ref{sec:workProcess}, the software development process should always be streamlined with the respective project requirements. 

While we focused on specific recommendations a single researcher can apply, additional topics came up in our workshops. 
The participants appreciated the exchange on \ac{RSE} with other researchers during the workshops. These exchanges can also be established within institutions, e.g., as a regular informal exchange round. 
Also, the topic of training came up in the workshops. In energy research, many researchers are not formally trained in software engineering. Training courses can be offered on many topics covered by the recommendations. 
They are beneficial for version control, testing, clean code, and basic architectural knowledge.
These trainings can be offered by experts in the same institution or external trainers, e.g., from the digital research academy\footnote{\url{https://digital-research.academy/}, last access 2025-01-02}.
We recommend elaborating at the beginning of a project on which training can be helpful to increase the project team's software engineering knowledge.

\section{Conclusion}
\label{sec:outlook}

While \ac{ERS} engineering is fundamental for energy research, research on \ac{ERS} is relatively new, and many aspects are still unknown.
As discussed in \autoref{sec:dis}, recommendations for \ac{RSE} depend on the type of research software and its development team. To explore these dependencies, a better understanding of the different types of \ac{ERS} is needed. Hasselbring et al. \cite{hasselbring_toward_2024} developed a general categorization of research software, which can serve as a starting point.

While our recommendations are generally similar to the existing recommendations in computational biology, they also differ in details. By developing recommendations in other research communities as well, it would be possible to join these recommendations to universal recommendations for general \ac{RSE}.

The developed recommendations focus on individual approaches to develop better \ac{ERS}. However, there are also systemic challenges to achieving this goal. 
For example, infrastructure for research software is required and currently missing, e.g., a registry for \ac{ERS} (as discussed in \cite{ferenz_towards_2023}) and required ontologies/standards. Good infrastructure can lower the barrier to better handle \ac{ERS}.

Another aspect discussed intensively at the workshops was that software is not sufficiently perceived as a scientific contribution. Instead, the software is often only a means to perform experiments and create results for publications. 
However, a well-written and documented library can be expected to have a large scientific impact because it is directly reusable and extendable for other researchers. In conclusion, we as a scientific community should value research software more instead of focusing too much on publications. 
Most of these challenges are similar for all types of research software, where Anzt et al. \cite{anzt2020environment} provided a good overview. Nevertheless, we see a domain-specific discussion of these aspects as an important step to improve the quality of \ac{ERS} and, therefore, the quality of energy research.
Overall, our recommendations serve as a starting point to better understanding \ac{ERS} and improve its engineering in energy research.


\begin{acronym}
    \acro{ERS}{Energy Research Software}
    \acro{RSE}{Research Software Engineering}
    \acro{FAIR}{Findable Accessible Interoperable Reusable}
    \acro{CIM}{Common Information Model}
    \acro{JOSS}{Journal of Open Source Software}
    \acro{FMI}{Functional Mock-up Interface}
    \acro{CFF}{Citation File Format}
    \acro{CICD}[CI/CD]{Continuous Integration/Continuous Delivery}
\end{acronym}
\begin{acks}
The authors would like to thank all participants of the two workshops for their time and valuable insights with special thanks to Anna-Lena Lamprecht for her wonderful talk on RSE. Also, the authors would like to thank the German Federal Government, the German State Governments, and the Joint Science Conference (GWK) for their funding and support as part of the NFDI4Energy and NFDI4Ing consortia. The work was partially funded by the German Research Foundation (DFG) – 501865131 and 442146713 within the German National Research Data Infrastructure (NFDI, www.nfdi.de) and 359941476 within the Schwerpunktprogramm 1984.
\end{acks}

\bibliographystyle{ACM-Reference-Format}
\bibliography{references}

\end{document}